\documentclass[11pt,a4,onecolumn,secnumarabic,amssymb, amsmath, nofootinbib,tightenlines,
nobibnotes, aps, prl,epsfig]{revtex4}
\usepackage{graphicx}
\usepackage{dcolumn}
\usepackage{bm}
\begin{document}
\preprint{APS/123-QED}
\title{Analytical approach for the approximation solution of the independent DGLAP evolution equations with respect to the hard Pomeron Behavior}

\author{B.Rezaei }
\altaffiliation{brezaei@razi.ac.ir}
\author{G.R.Boroun}
 \email{boroun@razi.ac.ir}
\affiliation{ Physics Department, Razi University, Kermanshah
67149, Iran}
\date{\today}

\begin{abstract}
We show that it is possible to use hard-Pomeron behavior to the
gluon distribution and singlet structure function at low $x$. We
derive a second-order independent differential equation for the
gluon distribution and the singlet structure function. In this
approach, both singlet quarks and gluons have the same high-energy
behavior at small $x$. These equations are derived from the
next-to-leading order DGLAP evolution equations. All results can
be consistently described in the framework of perturbative QCD,
which shows an increase of gluon distribution and singlet
structure functions as $x$ decreases.
\end{abstract}
\maketitle
The DGLAP $[1]$ evolution equations are fundamental tools to study
the $Q^{2}$ and $x$ evolutions of structure functions, where $x$
is the Bjorken scaling and and $Q^{2}$ is the four momenta
transfer in a deep inelastic scattering (DIS) process $[2]$. The
measurements of the $F_{2}(x,Q^{2})$ structure functions by DIS
processes in the small- $x$ region have opened a new era in parton
density measurements inside hadrons. The structure function
reflects the momentum distributions of the partons in the nucleon.
It is also important to know of the gluon distribution inside a
hadron at low- $x$ because gluons are expected to be dominant in
this region. The steep rise of $F_{2}(x,Q^{2})$ towards low $x$
observed at HERA also indicates in perturbative quantum
chromodynamics (PQCD) a similar rise of the gluon distribution
towards low $x$. In the usual procedure the DIS data are analyzed
by the NLO QCD fits based on the numerical solution of the DGLAP
evolution equations and it is found that the DGLAP analysis can
well describe the data in the perturbative region $Q^{2}{\geq}1
GeV^{2}$ [3]. As a alternative to the numerical solution, one can
study the behavior of the quarks and gluons through the analytical
solutions of the evolution equations. Although exact analytic
solutions of the DGLAP equations are not possible in the entire
range of $x$ and $Q^{2}$, such solutions are possible under
certain conditions [4-5] and are then quite then successful as far
as
the HERA small $x$ data are concerned.\\
Small $x$ behavior of structure functions for fixed $Q^{2}$
reflects the high energy behavior of the virtual Compton
scattering total cross section with increasing the total
center-of-mass energy squared $W^{2}$ becuse $W^{2}=Q^{2}(1/x-1)$.
The appropriate framework for the theoretical description of this
behavior is the Regge pole exchange picture [6]. It can be
asserted confidently that Regge theory is one of the most
successful approaches to the description of high energy scattering
of hadrons. This high energy behavior can be described by two
contributions: an effective Pomeron with its intercept slightly
above unity ($\sim$1.08) and the leading meson Regge trajectories
with intercept $\alpha_{R}(0){\approx}0.5$ [7]. The hypothesis of
the Pomeron with data of the total cross section shows that a
better description is achieved in alternative models with the
Pomeron having intercept one, but with a harder $j$ singularity (a
double pole) [8]. This model has two Pomeron components, each of
them with intercept $\alpha_{P}=1$; one is a double pole and the
other one is a simple pole [9]. It is tempting, however, to
explore the possibility of obtaining approximate analytical
solutions of DGLAP equations themselves in the restricted domain
of low- $x$ at least. Approximate solutions of DGLAP equations
have been reported $[10-12]$ with considerable phenomenological
success. In such an approximate scheme, one uses a Taylor
expansion valid at low- $x$ and reframes the DGLAP equations as
partial differential equations in the variable $x$ and $Q^{2}$
which can be solved by standard
methods.\\
In the past three decades, some authors were the first to report a
detailed look at the Regge input into the DGLAP equations [13-15].
So that we have shown [16-19] that it was possible to use Regge-
like behavior as an input for the
Dokshitzer-Gribov-Lipatov-Altarelli-parisi(DGLAP) evolution
equations at low $x$. The small $x$ region of the Deep inelastic
electron- proton scattering(DIS) offers a unique possibility to
explore the Regge limit of perturbative quantum
chromodynamic(PQCD)[6]. This model gives the following
parametrizations of the DIS distribution functions,
$f_{i}(x,Q^{2})=A_{i}(Q^{2})x^{-\lambda_{i}}$ (i=$\Sigma$(singlet
structure function) and g(gluon distribution)), where
$\lambda_{i}$ is Pomeron intercept minus one, show that
$\lambda_{i}=dlnf_{i}(x,Q^{2})/dln1/x$ definitely rises with
$Q^{2}$. In the present article we concentrate on the Regge
behavior in our calculations, although good fits to the results
clearly show that the gluon distribution and the singlet structure
function need a model with a hard Pomeron. In such scheme, one
uses this behavior that is valid at low $x$ and reframes the DGLAP
evolution equations as independent partial differential equations
in the variables $x$ and $Q^2$, which can be solved by standard
methods. Also, we should be able to calculate $\lambda_{s}$ and
$\lambda_{g}$ in the next- to-leading order(NLO) DGLAP equations.

 The NLO- DGLAP equations for the evolution of the
singlet structure function and the gluon distribution can be
written as
\begin{eqnarray}
\frac{dG(x,Q^{2})}{dlnQ^{2}}&=&\frac{\alpha_{s}}{2\pi}{\int_{0}^{1-x}}dz[P^{LO+NLO}_{gg}(1-z)G(\frac{x}{1-z},Q^{2})+P^{LO+NLO}_{gq}(1-z)\Sigma(\frac{x}{1-z},Q^{2})]\nonumber\\
\frac{d\Sigma(x,Q^{2})}{dlnQ^{2}}&=&\frac{\alpha_{s}}{2\pi}{\int_{0}^{1-x}}dz[P^{LO+NLO}_{qq}(1-z)\Sigma(\frac{x}{1-z},Q^{2})+2n_{f}P^{LO+NLO}_{qg}(1-z)G(\frac{x}{1-z},Q^{2})]\nonumber\\
\end{eqnarray}
Here $G(x,Q^{2})=xg(x,Q^{2})$ and
$\Sigma(x,Q^{2})=\frac{18}{5}F_{2}^{ep}(x,Q^{2})$ (at small $x$,
the nonsinglet contribution $F_{2}^{Ns}(x,Q^{2})$ is negligible
and can be ignored). In the evolution kernels and the running
coupling we take $N_{f}=4$(the number of active flavors), also for
simplicity we will ignore the threshold factors, which become
irrelevant for $Q^{2}>>4M_{i}^{2}$, and illustrate our method
using two quark families, $u,d,s,c$. Then
$n_{f}={\sum}e_{i}^{2}=\frac{10}{9}$. $P_{ij}$ are the NLO
splitting functions for quarks and gluons. The formal expressions
for these functions are fully known to NLO[20].\\
Let us first inserting the hard Pomeron behavior of the parton
distribution functions (PDFs) in the DGLAP evolution equations.
After integrating terms, rewrite Eqs.(1), as we find a set of
coupled formula to extract the gluon distribution and singlet
structure function.
\begin{eqnarray}
\frac{dG}{dt}&=&\frac{\alpha_{s}}{2\pi}[G(x,t)\eta_{1}+\Sigma(x,t)\beta_{1}]\nonumber\\
\frac{d\Sigma}{dt}&=&\frac{\alpha_{s}}{2\pi}[G(x,t)\eta_{2}+\Sigma(x,t)\beta_{2}]
\end{eqnarray}
where
\begin{eqnarray}
\eta_{1}&=&\frac{2C_{A}(1-x^{\lambda_{g}})+\frac{\alpha_{s}}{2\pi}(12C_{F}N_{f}T_{R}-46C_{A}N_{f}T_{R})(1-x^{\lambda_{g}})}{\lambda_{g}}\nonumber\\
\beta_{1}&=&\frac{2C_{F}(1-x^{\lambda_{s}})+\frac{\alpha_{s}}{2\pi}(9C_{F}C_{A}-40C_{F}N_{f}T_{R})(1-x^{\lambda_{s}})}{\lambda_{s}}\nonumber\\
\eta_{2}&=&\frac{\frac{\alpha_{s}}{2\pi}40C_{A}N_{f}T_{R}(1-x^{\lambda_{g}})}{\lambda_{g}}\nonumber\\
\beta_{2}&=&\frac{\frac{\alpha_{s}}{2\pi}40C_{F}N_{f}T_{R}(1-x^{\lambda_{s}})}{\lambda_{s}}.
\end{eqnarray}
For an SU(N) gauge group we have $C_{A}=N$, $C_{F}=(N^{2}-1)/2N$,
 $T_{F}=N_{f}T_{R}$, and $T_{R}=1/2$, that $C_{F}$ and $C_{A}$ are the color Cassimir operators.
The running coupling constant $\frac{\alpha_{s}}{2\pi}$ has the
form in the NLO as
\begin{equation}
\frac{\alpha_{s}}{2\pi}=\frac{2}{\beta_{0}t}[1-\frac{\beta_{1}lnt}{\beta_{0}^{2}t}]
\end{equation}
with $\beta_{0}=\frac{1}{3}(33-2N_{f})$ and
$\beta_{1}=102-\frac{38}{3}N_{f}$, also the variable $t$ is
defined by $t=ln(\frac{Q^{2}}{\Lambda^{2}})$ and the $\Lambda$ is
the QCD cut- off parameter.\\
Now, we combine terms and define relation between exponents of the
gluon and singlet distributions. According to the Regge theory,
the high energy (low $x$) behavior of both gluons and sea quarks
is controlled by the same singularity factor in the complex
angular momentum plane [6], and so we would expect
$\lambda_{s}=\lambda_{g}=\lambda$. We have fitted exponents to the
power law in low $x$ limit that we took for the PDFs. In Regge
theory the high energy behavior of hadron-hadron and photon-hadron
total cross section is determined by the pomeron intercept
$\alpha_{P}=1+\lambda$, and is given by
$\sigma_{\gamma(h)p}^{tot}(\nu){\sim}\nu^{\lambda}$. This behavior
is also valid for a virtual photon for $x<<1$, leading to the well
known behavior,$F_{2}{\sim}x^{-\lambda}$, of the structures at
fixed $Q^{2}$ and $x{\rightarrow}0$ [21-23]. The power $\lambda$
is found to be either $\lambda=0$ or $\lambda=0.5$. The first
value corresponds to the soft Pomeron and the second value the
hard (Lipatov) Pomeron intercept.   The Form $x^{-\lambda_{g}}$
for the gluon parametrization at small $x$ is suggested by Regge
behavior, but whereas the conventional Regge exchange is that of
the soft Pomeron, with $\lambda_{g}{\sim}0.0$, one may also allow
for a hard Pomeron with $\lambda_{g}{\sim}0.5$. The form
$x^{-\lambda_{s}}$ in the sea quark parametrization comes from
similar considerations since, at small $x$, the process
$g{\rightarrow}\hspace{0.1cm}q\overline{q}$ dominates the
evolution of the sea quarks. Hence the fits to early HERA data
have as a constraint $\lambda_{s}=\lambda_{g}=\lambda$, as the
value of $\lambda$ should be close to $0.5$ in quite a broad range
of low $x$ [4,7-9,24].\\
 After
successive differentiations of both sides of Eqs.(2),
multiplication by $G^{-1}(x,t)$, and some rearranging, we find an
inhomogenous second- order differential equation which determines
$\lambda_{g}$ and $\lambda_{s}$ in terms of $t$ independently, as
for $\lambda_{g}$  and $\lambda_{s}$ we find that

\begin{eqnarray}
\frac{2\pi}{\beta_{1}\alpha_{s}}
lnx\frac{d^{2}\lambda_{g}}{dt^{2}}-[\frac{\eta_{1}+\beta_{2}}{\beta_{1}}-2\pi\frac{d(\beta_{1}\alpha_{s})^{-1}}{dt}+\frac{2\pi}{\beta_{1}\alpha_{s}}
(lnx\frac{d\lambda_{g}}{dt})](lnx\frac{d\lambda_{g}}{dt})+\frac{d(\eta_{1}\beta_{1}^{-1})}{dt}-\frac{\alpha_{s}}{2\pi}(\frac{\eta_{1}
\beta_{2}}{\beta_{1}}-\eta_{2})=0\nonumber\\
\end{eqnarray}
and

\begin{eqnarray}
\frac{2\pi}{\eta_{2}\alpha_{s}}
lnx\frac{d^{2}\lambda_{s}}{dt^{2}}-[\frac{\eta_{1}+\beta_{2}}{\eta_{2}}-2\pi\frac{d(\eta_{2}\alpha_{s})^{-1}}{dt}+\frac{2\pi}{\eta_{2}\alpha_{s}}
(lnx\frac{d\lambda_{s}}{dt})](lnx\frac{d\lambda_{s}}{dt})+\frac{d(\beta_{2}\eta_{2}^{-1})}{dt}-\frac{\alpha_{s}}{2\pi}(\frac{\eta_{1}
\beta_{2}}{\eta_{2}}-\beta_{1})=0\nonumber\\
\end{eqnarray}
 The presented results gives the independent evolution equations
for the gluon and also singlet structure function exponents at
small $x$. These equations show that the exponents are functions
of $Q^{2}$. The $lnQ^{2}$ dependence of the exponents have a two-
order polynomial behavior. As, by solving these evolution
equations, we can determined exponents with the starting
parameterizations of exponents
($\lambda_{i}(t_{0})=dlnf_{i}(x,t_{0})/dln1/x$) given by the input
distributions of the partons and its derivatives, respectively
[25,27-29]. Therefore, the effective power- law behavior of the
gluon distribution and the singlet structure function corresponds
to:
\begin{equation}
f_{i}(x,t)=f_{i}(x,t_{0})x^{-(\lambda_{i}(t)-\lambda_{i}(t_{0}))},(i=\Sigma,
g)
\end{equation}
 If we want to perform parton distribution functions, we need to
fix these at an initial scale
$t_{0}=\ln\frac{Q_{0}^{2}}{\Lambda^{2}}$. Here we used the QCD
cut- off parameter
$\Lambda^{4}_{\overline{MS}}=0.323\hspace{0.1cm}GeV$
 [11] for $\alpha_{s}(M_{z^{2}})=0.119$. Also in our calculations,
 we need the initial conditions $f_{i}(x,t_{0})$ and
 $\lambda_{i}(t_{0})$ that are according to the input parameterization. In order to test the
validity of our obtained gluon distribution, we calculate the
gluon (or singlet) distribution functions and exponent of the
gluon (or singlet) distribution using Eq.(7) and compare them with
the theoretical predictions starting with the evolution at
$Q_{0}^{2}=5 \hspace{0.1cm}$GeV$^{2}$. The results of calculation
are shown in Figs.1 and 2 at several $Q^{2}$ values. As it can be
seen in these figures,
  the values of $f_{i}(x,Q^{2})$ increase as
  $x$ decreases. In these figures We compare our results for the gluon distribution function and also the proton structure function
   with the DL fit [7,29] and H1 data [27] with the total errors at $Q^{2}$
   values. Also, we compared our predictions for the proton
   structure function with Ref.[23]. As the proton structure
   function is corresponding to the gluon distribution function at
   $x_{g}{\approx}2x$ upon integration form the DGLAP equation. In
   this integration we used from the standard GRV parameterizations. We have taken
the DL parametric form for the starting distribution  at
$Q_{0}^{2}=5 GeV^{2}$  given by
$xg(x,Q^{2})=0.95(Q^{2})^{1+\epsilon_{0}}(1+Q^{2}/0.5)^{-1-\epsilon_{0}/2}x^{-\epsilon_{0}}$
where $\epsilon_{0}$ is equal to $0.437$ according to hard Pomeron
exchange. We can observe that
  these distribution function values increases when $x$ decreases, but with a somewhat smaller
rate. This behavior is associated with the exchange of an object
known as the hard Pomeron. Having concluded that the data for
$F_{2}$ require a hard pomeron component, it is necessary to test
this with our results, as compared in Fig.2.\\

To conclude, in this paper we have obtained independently
solutions for the gluon and singlet exponents based on the DGLAP
evolution equations with respect to the Regge behavior in the
next- to- leading order (NLO) at low $x$. Careful investigation of
our results shows a agreement with the previously published parton
distributions based on QCD. The gluon distribution and singlet
structure functions  increase as usual, as $x$ decreases. The form
of the obtained distribution functions
 for the gluon distribution and the singlet structure functions  are similar to the one predicted
 from the parton parameterization. The formulas  used to generate
 the parton distributions are in agreement with the increase observed by
 H1 experiments. Also these results show that
 the exponents increases non linearly
 with respect to ${\ln}Q^{2}$ as $x$ decreases. So that, the behaviors of the distribution
 functions at low $x$ are consistent
 with a power- law behavior. The obtained results give strong indications that the proposed
formulae, being very simple, provides relatively accurate values
for the gluon distribution and structure function.\\


\textbf{References}\\
1. Yu. L.Dokshitzer, Sov.Phys.JETPG {\bf6}, 641(1977 );
G.Altarelli and
G.Parisi, Nucl.Phys.B{\bf126}, 298(1997 ); V.N.Gribov and L.N.Lipatov, Sov.J.Nucl.Phys.{\bf28}, 822(1978).\\
2. L.F.Abbott, W.B.Atwood and A.M.Barnett, Phys.Rev.D {\bf22}, 582(1980).\\
3. A.M.Cooper- Sarkar, R.C.E.Devenish and A.DeRoeck, Int.J.Mod.Phys.A {\bf13}, 3385( 1998 ).\\
4. A.K.Kotikov  and G.Parente, Phys.Lett.B {\bf379}, 195(1996 ); J.Kwiecinski, hep-ph/9607221.\\
5. R.D.Ball and  S.Forte, Phys.Lett.B {\bf335},  77(1994 )\\; Phys.Lett.B {\bf336}, 77(1994 ).\\
6.  P.D.Collins, An introduction to Regge theory and
high-energy physics(Cambridge University Press,1997).\\
7. A.Donnachie and  P.V.Landshoff, Phys.Lett.B {\bf296}, 257(1992).\\
8. P.Desgrolard, M.Giffon, E.Martynov and E.Predazzi, Eur.Phys.J.C {\bf18}, 555(2001).\\
9. P.Desgrolard, M.Giffon and E.Martynov, Eur.Phys.J.C {\bf7},  655(1999).\\
10. M.B.Gay Ducati  and V.P.B.Goncalves, Phys.Lett.B {\bf390}, 401(1997).\\
11. K.Pretz, Phys.Lett.B {\bf311}, 286(1993); Phys.Lett.B {\bf332}, 393(1994).\\
12. A.V.Kotikov, hep-ph/9507320.\\
13. C.Lopez and F.J.Yndurain, Nucl.Phys.B{\bf171}, 231(1980).\\
14. C.Lopez, F.Barreiro and F.J.Yndurain, Z.Phys.C{\bf72}, 561(1996).\\
15. K.Adel, F.Barreiro and F.J.Yndurain, Nucl.Phys.B{\bf495}, 221(1997).\\
16. G.Soyez, Phys.Rev.D\textbf{67}, 076001(2003).\\
17. L.Csernai, et.al, Eur.Phys.J.C\textbf{24}, 205(2002).\\
18. G.Soyez, Phys.Rev.D\textbf{71}, 076001(2005).\\
19. G.R.Boroun and B.Rezaie, Phys.Atom.Nucl.\textbf{71}, No.6,
 1076(2008); G.R.Boroun, JETP,\textbf{133}, No.4, 805(2008).\\
20. R.K.Ellis , W.J.Stirling and B.R.Webber, QCD and Collider
Physics(Cambridge University Press,1996).\\
21. N.Nikolaev, J.Speth and V.R.Zoller,Phys.Lett.B{\bf473}, 157(2000).\\
22. R.Fiore, N.Nikolaev and V.R.Zoller,JETP Lett{\bf90}, 319(2009).\\
23. I.P.Ivanov and N.Nikolaev,Phys.Rev.D{\bf65},054004(2002).\\
24. A.D.Martin, M.G.Ryskin and G.Watt,
arXiv:hep-ph/0406225(2004).\\
25. C.Adloff et.al, $H{1}$ Collab., Eur.Phys.J.C\textbf{21}, 33(2001); Phys.Lett.B\textbf{520}, 183(2001)\\
26. A.M.Cooper- Sarkar and R.C.E.D Evenish, Acta.Phys.Polon.B \textbf{34}, 2911(2003).\\
27. C.Adloff, et.al, $H1$ Collab. Eur.Phys.J.C {\bf21}, 33(2001 ).\\
28. M.Gluk, E.Reya  and A.Vogt, Z.Phys.C {\bf67}, 433(1995 ); Euro.Phys.J.C {\bf5}, 461(1998 ).\\
29.P.V.Landshoff, hep-ph/0203084.

\begin{figure}
\includegraphics[width=1\textwidth]{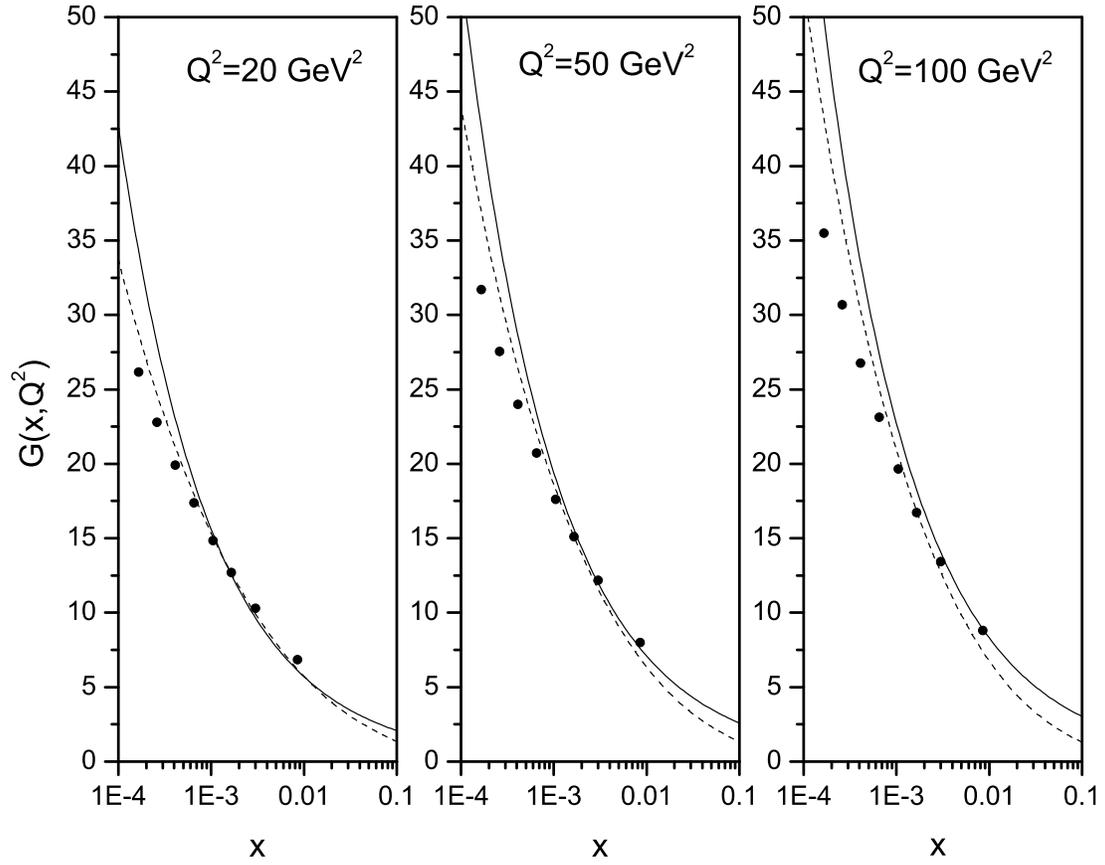}
\caption{The gluon distribution against $x$ at fixed $Q^{2}$
values (solid circles) compared with the DL fit [7,29](solid line)
and the GRV parameterization[28] (dashed line).}
\end{figure}
\begin{figure}
\includegraphics[width=1\textwidth]{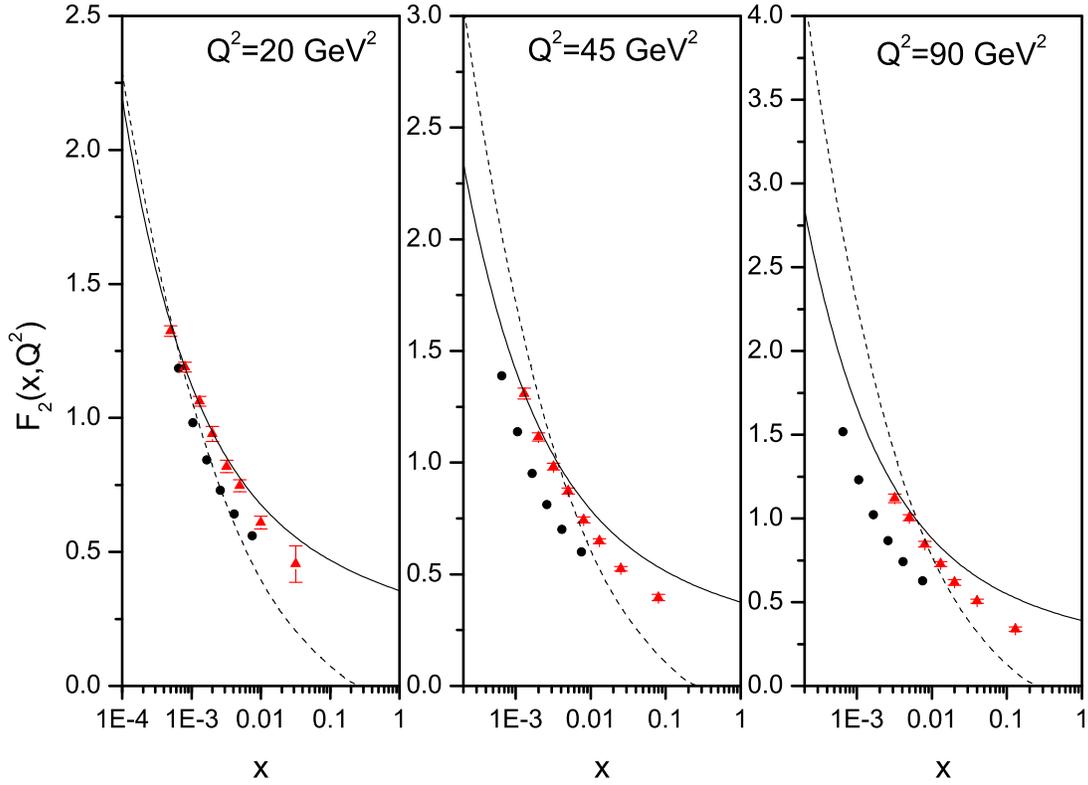}
\caption{The calculated values of the structure function $F_{2}$
for several values of $Q^{2}$ plotted as function of $x$ with the
starting parameterization of the structure function at
$Q_{0}^{2}=5 GeV^{2}$ (solid circles), compared with the next- to-
leading order QCD fit to the H1 data with total errors (up
triangle) also with the DL fit[7,29] (solid line) and dotted line
represents Ref.[23] results for the GRV parameterization of the
gluon distribution function to the proton structure function.}
\end{figure}

\end{document}